\documentclass[12pt]{article}
\newcommand\HH{{\cal H}}
\newcommand\LL{{\cal L}}

\newcommand\Ol{{\cal O}_\lambda}
\newcommand\h{\hbar}
\newcommand{\p}{\partial}
\newcommand\ind{{\rm ind}\,}
\newcommand\G{{\cal G}}

\newcommand\la{\lambda}
\newcommand\eq{\eqno}
\newcommand{\lug}{\langle}
\newcommand{\rug}{\rangle}
\newcommand\rank{{\rm rank}\,}
\newcommand\corank{{\rm corank}\,}
\newcommand\s{{(s)}}
\newcommand\rs{{r_{(s)}}}
\newcommand\os{\omega^{(s)}}
\newcommand\Os{{\Omega^{(s)}}}

\newcommand\be{\begin{eqnarray*}}
\newcommand\ee{\end{eqnarray*}}
\newcommand\Fsa{F^{(s)}_\alpha}
\newcommand\Ksm{K^{(s)}_\mu}
\newcommand\al{\alpha}
\newcommand\LLs{\LL_{(s)}}
\newcommand\vp{\varphi}
\newcommand\tl{\tilde}

\author{I. V. Shirokov}
\title{
DARBOUX COORDINATES ON K-ORBITS AND THE SPECTRA OF
CASIMIR OPERATORS ON LIE GROUPS
}
\date{}

\begin{document}


\maketitle

\begin{abstract}
We propose an algorithm for obtaining the spectra of Casimir (Laplace)
operators on Lie groups. We prove that the existence of the normal
polarization associated with a linear functional on the Lie algebra is
necessary and sufficient for the transition to local canonical Darboux
coordinates $(p,q)$ on the coadjoint representation orbit that is linear in
the "momenta." We show that the $\la$-representations of Lie algebras (which
are used, in particular, in integrating differential equations) result from
the quantization of the Poisson bracket on the coalgebra in canonical
coordinates.
\end{abstract}

\section*{Introduction}

The method of orbits discovered in the pioneering
works of Kirillov [1, 2] (see also [3, 4]) is a universal base for
performing harmonic analysis on homogeneous spaces and for constructing new
methods of integrating linear differential equations. Steps towards
implementing this program were described in [5]. The main result of this
work is an algorithm for obtaining the spectra of Casimir (Laplace)
operators on a Lie group via linear algebraic methods starting with the
known structure constants and the information regarding the compactness of
certain subgroups. We show that the existence of a normal polarization
associated with a linear functional $\la$ is necessary and sufficient
for the existence of local canonical Darboux coordinates $(p, q)$ on the
K-orbit $\Ol$ such that the transition to these coordinates is linear in the
"momenta." (We developed a computer program based on the computer algebra
system Maple V to evaluate the canonical Darboux coordinates for a given
functional and the corresponding normal polarization.) The subsequent
"quantization" leads to the notion of $\la$-representations of Lie algebras
consisting in assigning each K-orbit a special representation of the Lie
algebra via differential operators. The $\la$-representations first appeared
in the noncommutative integration method of linear differential equations
[6] as a "quantum" analogue of the noncommutative Mishchenko-Fomenko
integration method for finite-dimensional Hamiltonian systems [7]. The
$\la$-representation operators are also implicitly involved as the generators
of irreducible representations of Lie groups.

\section{The description of K-orbits}

Let $G$ be a real connected $n$-dimensional Lie group and $\G$ be its Lie
algebra. The action of the adjoint representation $Ad^*$ of the Lie group
defines a fibration of the dual space $\G^*$ into even-dimensional
orbits (the K-orbits). The maximum dimension of a K-orbit is $n - r$,
where $r$ is the index ($\ind\G$) of the Lie algebra defined as the
dimension of the annihilator of a general covector. We say that a linear
functional (a covector) $\la$ has the degeneration degree $s$ if it belongs
to a K-orbit $\Ol$ of the dimension $\dim \Ol = n-r-2s, s = 0.... ,(n-r)/2.$

We decompose the space $\G^*$ into a sum
of nonintersecting invariant algebraic surfaces $M_s$ consisting of
K-orbits with the same dimension. This can be done as follows.
We let $f_i$ denote the coordinates of the covector $f$ in the dual basis,
$f=f_ie^i$ with $\lug e^i,e_j\rug =\delta_j^i$, where
$\{e_j\}$ is the basis of $\G$. The vector fields on $\G^*$
$$
Y_i(f)\equiv C_{ij}(f)\frac{\p}{\p f_j},\quad C_{ij}(f)\equiv C_{ij}^kf_k
$$
are generators of the transformation group $G$ acting on the space
$\G^*$, and their linear span therefore
constitutes the space $T_f\Ol$ tangent to the orbit $\Ol$ running through
the point $f$. Thus, the dimension of the orbit $\Ol$ is determined by the
rank of the matrix $C_{ij}$,
$$
\dim \Ol=\rank C_{ij}(\la).
$$
It can be easily verified that the
rank of $C_{ij}$ is constant over the orbit. Therefore, equating the
corresponding minors of $C_{ij}(f)$ to zero and "forbidding" the vanishing
of lower-order minors, we obtain polynomial equations that define a surface
$M_s$,
\begin{eqnarray*}
M_0&=&\{f\in \G^*\mid \neg(F^{1}(f)=0)\};\\
M_s&=&\{f\in \G^*\mid F^{s}(f)= 0,\ \neg(F^{s+1}(f)= 0)\},\ s=1,\ldots,
\frac{n-r}{2}-1;\\
M_\frac{n-r}2 &=&\{f\in \G^*\mid F^{\frac{n-r}2}(f)= 0\}.
\end{eqnarray*}
Here, we let $F^s(f)$ denote the
collection of all minors of $C_{ij}(f)$ of the size $n - r - 2s + 2$,
the condition $F^s(f) = 0$ indicates that all the minors of $C_{ij}(f)$
of the size $n- r- 2s +2$
vanish at the point $f$, and $\neg(F^{s}(f)=0)$ means that the corresponding
minors do not vanish simultaneously at $f$.

The space $M_s$ can also be defined
as the set of points $f$ where all the polyvectors of degree $n- r-2s+1$
of the form $Y_{i_1}(f)\wedge\dots\wedge Y_{i_{n-r-2s+1}}(f)$
vanish, but not all the polyvectors of degree $n- r- 2s -1$ vanish.

We note that in the general case, the surface $M_s$
consists of several nonintersecting invariant components, which we
distinguish with subscripts as $M_s=M_{sa}\cup M_{sb}\ldots.$
(To avoid stipulating
each time that the space $M_s$ is not connected, we assume the convention
that $s$ in parentheses, $(s)$, denotes a specific type of the orbit
with the degeneration degree s.) Each component $M_\s$ is defined by the
corresponding set of homogeneous polynomials $F^\s_\alpha(f)$
satisfying the conditions
$$
Y_i\Fsa(f)\bigr|_{F^\s(f)=0}=0. \eq (1.1)
$$
Although the invariant algebraic surfaces $M_\s$ are not linear spaces,
they are star sets, i.e., $f\in M_\s$, implies $tf\in M_\s$ for  $t\in R^1$.

{\bf Example 1} (The Poincare group $P^{1,3}$).
The Poincare group $P^{1,3}=T^4\triangleright SO(1,3)$ is the group of
motions in Minkowski space and is a semidirect product of the semisimple
Lorentz rotation group $SO(1,3)$ with the four-dimensional commutative group of
translations $T^4$. The commutation relations of the Poincare algebra
${\cal P}^{1,3}=\{e_{ab},e_a\}$ can be written as
\begin{eqnarray*}
&&[e_{ab},e_{cd}]=g_{ad}e_{bc}-g_{ac}e_{bd}+g_{bc}e_{ad}-g_{bd}e_{ac};
\quad [e_a,e_{bc}]=g_{ab}e_c-g_{ac}e_b;\\
&&[e_a,e_b]=0;\quad a,b,c,d=0,1,2,3;\quad g_{ab}={\rm diag}(1,-1,-1,-1).
\end{eqnarray*}
We have $n = 10$ and $r = 2$ in this example. With $\{l_{ab},p_a\}$
denoting the coordinates of a covector $f$ in the dual basis, we give the
explicit form of the surfaces $M_\s$:
$$
({\cal P}^{1,3})^*=M_0 \bigcup M_{1a} \bigcup M_{1b} \bigcup M_2
\bigcup M_4,
$$
\be
M_0&=&\{f\in R^{10} \mid \neg(W_ap_b-W_bp_a=0)\};\ \dim M_0=10;\\
M_{1a}&=&\{f\in R^{10}\mid W_a=0,\ \neg(p_a=0)\};\ \dim M_{1a}=7;\\
M_{1b}&=&\{f\in R^{10}\mid W_ap_0=W_0p_a,\ \neg(W_a=0)\};\ \dim M_{1b}=7;\\
M_2&=&\{f\in R^{10}\mid p_a=0, \ \neg(f=0)\};\ \dim M_2=6\\
M_3&=&\emptyset;\qquad M_4=\{f=0\};\quad \dim M_4=0.
\ee
(We introduce the notation $\ W^a\equiv \frac12 \varepsilon^{abcd}l_{bc}p_d$,
and the indices are raised and lowered by the diagonal
matrix $g_{ab}$.) We note that in view of the identity $W^ap_a = 0$,
there are only
three independent functions among the four functions $W^a$.

The dual space ${\cal G}^*$
has a degenerate linear Poisson bracket
$$
\{\varphi,\psi\}(f)\equiv \lug f,[\nabla \vp (f),\nabla \psi (f)]\rug;
\quad  \varphi,\psi \in C^\infty({\cal G}^*). \eq (1.2)
$$
The functions $K^\s_\mu(f)$ that are nonconstant on $M_\s$ are called the
{\it $\s$-type Casimir functions} if they commute with any function on
$M_\s$.

The $\s$-type Casimir functions can be found from the equations
$$
C_{ij}(f)\frac{\p K_\mu^\s(f)}{\p f_j}\biggr|_{f\in M_\s}=0,\quad i=1,
\dots,n. \eq (1.3)
$$
It is obvious that the
number $r_\s$ of independent $(s)$-type Casimir functions is related to the
dimension of the space $M_\s$ as $\rs=\dim M_\s + 2s+r-n$. Because $M_\s$
are star spaces, we can assume without loss of generality that the Casimir
functions $K^\s_\mu(f)$ are homogeneous,
$$
\frac{\p K_\mu^\s(f)}{\p f_i}f_i=m_\mu^\s K_\mu^\s(f) \Longleftrightarrow
K^\s_\mu(tf)=t^{m^\s_\mu}K^\s_\mu(f);\quad \mu=1,\dots,\rs.
$$

In the general case, the Casimir functions are multivalued (for example, if
the orbit space $\G^*/G$ is not semiseparable, the Casimir functions are
infinitely valued), hi what follows, we use the term "Casimir function" to
mean a certain fixed branch of the multivalued function $K^\s_\mu$.
In the general case, the Casimir functions $K^\s_\mu$ are only locally
invariant under the coadjoint representation, i.e., the equality
$K^\s_\mu(Ad^*_gf)=K^\s_\mu(f)$ holds for
the elements $g$ belonging to some neighborhood of unity in the group $G$.

{\bf Remark 1}. Without going into detail, we note that the spaces $M_\s$
are critical surfaces for some polynomial $(s-1)$-type Casimir functions,
which gives a simple and efficient way to construct the functions $F^\s$.

We now let $\Os\subset R^{r_\s}$ denote the set of values of the mapping
$K^\s:M_\s\to R^{r_\s}$ and
introduce a locally invariant subset $O_\os$ as the level surface,
$$
O_\os=\{f\in M_\s\mid K_\mu^\s(f)=\os_\mu,\ \mu=1,\dots,\rs;\ \os\in \Os\}.
$$
The dimension of $O_\os$ is the same as the dimension of the orbit
$\Ol\in M_\s$, where $\os=K^\s(\la)$. If the Casimir functions are single
valued, the orbit space in separable, and the set $O_\os$ then consists of
a denumerable (typically, finite) number or orbits; accordingly, we call
this level surface the {\it class of orbits}.

The space $\G^*$ thus consists of the union of connected invariant
nonintersecting algebraic surfaces $M_\s$; these, in turn, are union of the
classes of orbits $O_\os$:
$$
\G^*=\bigcup_\s M_\s=\bigcup_\s\bigcup_{\os\in \Os}O_\os. \eq (1.4)
$$

{\bf Example 2} (The Poincare group $P^{1,3}$, continuation of Example 1).
In this case, decomposition (1.4) (with $\Delta_{1,3}\equiv p_ap^a$)
becomes
\be
({\cal P}^{1,3})^*&=&(\bigcup_{\omega\in R^2}O^0_\omega)
\bigcup(\bigcup_{\omega\in R^1}O^{1a}_\omega)
\bigcup(\bigcup_{\omega\in \{0\}\setminus R^1}O^{1b}_\omega)\bigcup
(\bigcup_{\omega\in R^2}O^2_\omega)\bigcup \{0\},
\ee
where
\be
O^0_\omega&=&\{f\in M_0\mid W^aW_a=\omega^0_1,\ \Delta_{1,3}=\omega^0_2;\}
,\quad \Omega^0=R^2;\\
O^{1a}_\omega&=&\{f\in M_{1a}\mid \Delta_{1,3}=\omega^{1a}_1\},\quad
\Omega^{1a}=R^1;\\
O^{1b}_\omega&=&\{f\in M_{1b}\mid  \frac{W_0}{p_0}=\omega^{1b}_1\},
\quad \Omega^{1b}=\{0\}\setminus R^1;\\
O^2_\omega&=&\{f\in M_2\mid  l_{ab}l^{ab}=\omega_1^2,\
\varepsilon^{abcd}l_{ab}l_{cd}= \omega_2^2;\
\},\quad \Omega^2=R^2.
\ee
In this example, each class consists of several K-orbits (with the number
ranging from one to four depending on the signs of the $\os$ parameters).

We now consider the quotient space $B_\s=M_\s/G$, whose points are the
orbits $\Ol\in M_\s$. It is obvious that $\dim B_\s=\rs$. We introduce
local coordinates $j$ on $B_\s$. For this, we parameterize an $\s$-covector
$\la \in M_\s$ by real-valued parameters $j=(j_1,\dots,j_\rs)$, assuming
that $\la$ depends linearly on $j$ (this can be done because $M_\s$ is a
star surface):
$$
\la=\la(j)
$$
with
$$
F_\alpha^\s(\la(j))\equiv 0,\quad K^\s_\mu(\la(j))
=\os_\mu(j),\quad \det \frac{\p \os_\mu(j)}{\p j_\nu}\neq 0.
$$
In other words, $\la(j)$ is a local section of the bundle $M_\s\to B_\s$.
We let $\Theta_\s\equiv (\os)^{-1}(\Os)\subset R^\rs$ denote the inverse
image and $\Gamma^\s$ denote the discrete group of transformations of the
set $\Theta_\s$: $j\to \hat j$ such that $\os_\mu(\hat j)=\os_\mu(j)$.
Each point $j$ from the domain $J^\s\equiv \Theta_\s/\Gamma^\s\subset R^\rs$
then corresponds to a single class $O_\os$.

Elementary examples show that global parameterization does not exist on the
whole of $M_\s$ in general, i.e., the manifold $B_\s$ is not covered by one
chart. In this case, we define an atlas of charts on $B_\s$
and parameterize the
corresponding connected invariant nonintersecting subsets
$M_\s^A, M_\s^B,\dots$ with a
nonvanishing measure in $M_\s$ as follows:
$$
M_\s=M_\s^A\cup M_\s^B\cup\dots. \eq (1.5)
$$
The corresponding domains of values $J^A,J^B,\dots$ of the $j$ parameters
then satisfy the relation $\Os=\os(J^A)\cup\os(J^B)\cup \dots$.

In what follows, we show that each
space $M_\s^A$ corresponds to its own type of the spectrum of Casimir
operators on the Lie group. Therefore, we can say that decomposition (1.5)
is the decomposition with respect to the {\it spectral type}
and an $(s)$-orbit $\Ol$ belongs
to spectral type $A$ if $\Ol\in M_\s^A$. We illustrate decomposition (1.5)
with a simple example.

{\bf Example 3} (The group $SO(2,1)$). In the case of the group
$SO(2,1)$, $[e_1,e_2]=e_2,\ [e_2,e_3]=2e_1$, and $[e_3,e_1]=e_3$.
Decomposition (1.4) becomes
$$
O^0_\omega=\{f_1^2+f_2f_3=\omega,\ \neg(f=0)\},\quad O^1=\{f=0\}.
$$
For $\omega > 0$, the
class $O^0_\omega$ consists of two orbits. For nondegenerate orbits,
$\Omega=R^1$. There is
no single parameterization in this case. Indeed, the most general form of
the parameterization $\la(j) = (a_1j,a_2j,a_3j)$
(where $a_i$, are some numbers)
leads to $\omega(j)=aj^2$, where $a=a_1^2+a_2a_3$, and therefore
(depending on the sign of $a$) $\omega(j)$ is always greater than zero,
less than zero, or equal to zero, i.e., $\omega(R^1)\neq \Omega$.
We introduce two spectral types,
\be
{\rm type}\: A&:&\la(j)=(0,j,j);\quad J^A=[0,\infty);\quad O^{0A}_{\omega(j)}=
\{f_1^2+f_2f_3=j^2,\ f\neq 0\};\\
{\rm type}\: B&:&\la(j)=(0,j,-j);\quad J^B=(0,\infty);\quad O^{0B}_{\omega(j)}=
\{f_1^2+f_2f_3=-j^2,\ f\neq 0\}.
\ee
In what follows, we return to this example
and show that types $A$ and $B$ respectively correspond to the
continuous and the discrete spectra of the Laplace (Casimir) operator
on the group $SO(2,1)$.

The classification of K-orbits obtained above allows describing
the structure of the annihilator $\G^\la$ for an arbitrary (in general,
degenerate) covector $\la$ in more detail. Let $\la$ be a covector of
the $(s)$-type.
It follows from the definition of the annihilator that
$\dim\G^\la=\corank C_{ij}(\la)={\rm codim}\, \Ol=2s+r$.

It can be verified that the functions
$\Phi^\s_a(f)=(F^\s_\alpha(f),K^\s_\mu(f)), \;(a=1,\dots,2s+r)$,
have the Poisson brackets
\be
\{\Fsa,F^\s_\beta\}(f)&=&C_{\alpha\beta}^\gamma(f)F^\s_\gamma(f);\\
\{\Fsa,\Ksm\}(f)&=&C_{\alpha\mu}^\beta(f)F^\s_\beta(f);\\
\{\Ksm,K^\s_\nu\}(f)&=&C_{\mu\nu}^\alpha(f)\Fsa(f).
\ee
Because the functions $\Phi^\s_a(f)$ are independent, recalling the
definition of the space $M_\s$,
we conclude that the gradients $\nabla \Phi^\s(\la)$ are linearly
independent and constitute a basis of a $(2s+r)$-dimensional Lie
algebra $\G^\la$ with the
commutation relations
$$
[\nabla F_\alpha(\la),\nabla F_\beta(\la)]= C_{\alpha\beta}^\gamma(\la)
\nabla F_\gamma(\la);
$$
$$
[\nabla F_\alpha(\la),\nabla K_\mu(\la)]= C_{\alpha\mu}^\beta(\la)
\nabla F_\beta(\la); \eq (1.6)
$$
$$
[\nabla K_\mu(\la),\nabla K_\nu(\la)]= C_{\mu\nu}^\alpha(\la)
\nabla F_\alpha(\la);
$$
(where we omit the superscript $(s)$ for brevity). From these commutation
relations, we now have the following proposition.

{\bf Proposition 1}. {\it The
annihilator $\G^\la$ of an arbitrary $(s)$-covector $\la$ contains
the ideal ${\cal N}_\la=\{\nabla \Fsa(\la)\}$.
The quotient algebra
${\cal K}_\la=\G^\la/{\cal N}_\la=\{\nabla \Ksm(\la)+{\cal N}_\la\}$ is
commutative and $\rs$- dimensional}.

Closed subgroups of $G$ corresponding to the Lie
algebras $\G^\la$, ${\cal N}_\la$ and ${\cal K}_\la$
are respectively denoted by $G^\la$, $N_\la$ and $K_\la=G^\la/N_\la$.

\section{Darboux coordinates and $\la$-representations of Lie algebras}

We let $\omega_\la$ denote the Kirillov form on the orbit $\Ol$.
It defines a symplectic structure and acts on the vectors $a$ and $b$
tangent to the orbit as
$$
\omega_\la(a,b)=\lug \la,[\al,\beta]\rug,
$$
where $a=ad^*_\al \la$ and $b=ad^*_\beta \la$. The restriction of Poisson
brackets (1.2) to the orbit coincides with the Poisson bracket generated by
the symplectic form $\omega_\la$. According to the well-known Darboux
theorem, there exist local canonical coordinates (Darboux coordinates)
on the orbit $\Ol$ such that the form $\omega_\la$ becomes
$$
\omega_\la=dp_a\wedge dq^a;\quad a=1,\dots,\frac12\dim O_\la=\frac{n-r}{2}-s,
$$
where $s$ is the degeneration degree of the orbit.

Let $\la$ be an $(s)$-type covector and $f\in\Ol$. It can be easily seen
that the transition to canonical Darboux coordinates $(f_i)\to (p_a,q^a)$
amounts to constructing analytic functions $f_i=f_i(q,p,\la)$ of the
variables $(p,q)$ satisfying the conditions
$$
f_i(0,0,\la)=\la_i; \eq (2.1)
$$
$$
\frac{\p f_i(q,p,\la)}{\p p_a} \frac{\p f_j(q,p,\la)}{\p q^a}-
\frac{\p f_j(q,p,\la)}{\p p_a} \frac{\p f_i(q,p,\la)}{\p q^a}=
C_{ij}^kf_k(q,p,\la); \eq (2.2)
$$
$$
\Fsa(f(q,p,\la))=0, \quad
\Ksm(f(q,p,\la))= \Ksm(\la). \eq (2.3)
$$

We require that the transition to the canonical
coordinates (in other words, the $gp$-transition) be linear in $p_a$,
$$
f_i(q,p,\la)=\al^a_i(q)p_a+\chi_i(q,\la); \quad
\rank \al^a_i(q)=\frac12\dim \Ol. \eq (2.4)
$$
Obviously, a transition of form (2.4) does not exist in the general case;
however, assuming that $\al_i^a(q)$ and $\chi_i(q;\la)$ are holomorphic
functions of the complex variables $q$, we considerably broaden the class
of Lie algebras and
K-orbits for which this transition does exist. (We assume that functionals
from $\G^*$ can be continued to $\G^c$ by linearity.) It seems that
transition (2.4) exists for an arbitrary Lie algebra and any of its
nondegenerate orbits.

{\bf Theorem 1.} {\it The linear transition to canonica coordinates on the
orbit $\Ol$ exists if and only if there exists a normal polarization (in
general, complex) associated with the linear functional $\la$, i.e., a
subalgebra $\HH\subset \G^c$ such that}
$$
\dim \HH =n-\frac12 \dim \Ol,\quad \lug \la,[\HH,\HH]\rug=0,\quad
\la +\HH^\bot \subset \Ol.
$$

Before discussing and proving this theorem, we digress into a subject that
appears to be of independent interest. Let $X_i(x)=X_i^a(x)\p_{x^a}$ be
transformation group generators that generate an $n$-dimensional Lie algebra
$\G$ of vector fields on a homogeneous space
$M = G/H:\ [X_i,X_j] = C^k_{ij}X_k$ (here
and in what follows, $x^a\; (a=1,\dots,m=\dim M)$, are local coordinates
of a point $x\in M$); $H$ is the isotropy group of a base point $x_0$ and
$\HH$ is its Lie algebra. Inhomogeneous first-order operators
$\tl X_i=X_i+\chi_i(x)$ are called the
{\it continuations} of the generators $X_i$ if they still satisfy the
commutation relations of the algebra $\G$ ($\chi_i(x)$ and are smooth
functions on $M$).

By definition, the $n$-component function $\chi(x)$ is to be found from
the system of equations
$$
X_i^a(x)\frac{\p \chi_j(x)}{\p x^a}-X_j^a(x)\frac{\p \chi_i(x)}{\p x^a}=
C_{ij}^k\chi_k(x). \eq(2.5)
$$
Solutions of this system span
a linear space, in which we can single out the module of trivial solutions
of the form
$$
\chi^0=\left\{\chi^0_i(x)=X_i^a(x) \frac{\p S(x)}{\p x^a}\right\},
$$
where S(x) is an arbitrary
smooth function. Constructing trivial continuations
$\tl X_i=e^{-S}X_ie^S$ is
equivalent to performing the "gauge" transformation
$\p_{x^a}\to \p_{x^a}+\p_{x^a}S(x)$.
In what follows, we are interested in only nontrivial continuations that
generate the quotient space of all solutions of system (2.5) modulo trivial
solutions.

{\bf Proposition 2.} {\it The space of nontrivial continuations is finite
dimensional and is isomorphic to the quotient space} $\HH^*/[\HH,\HH]^*$.

We do not give the complete proof here; however, we give the explicit
form of all nontrivial solutions of system (2.5), which implies the
validity of Proposition 2. We relabel and change the basis in $\G$,
$$
X_a(x_0)= \frac{\p}{\p x^a}\bigg|_{x_0},\ a=1,\dots,m;\quad
X_\al(x_0)=0,\ \al =m+1,\dots,n.
$$
We consider the right action of $G$ on $M$. An
arbitrary element $g\in G$ is represented as $g = hs(x)$, where
$h\in  H$ and $s(x)$
is a smooth Borel mapping $M\to G$ assigning the right coset class
$Hs(x)\subset G$ to each point $x\in M$. In the coordinates
$g=(h^\al,x^a)$ (assuming that $e=(0,x_0)$),
the left-invariant vector fields $\xi_i(g)$ have the form
$\xi_i(g)=X_i^a(x)\p_{x^a}+\xi_i^\al(h,x)\p_{h^\al}$.
Direct calculation verifies that the continuations given by
the operators
$$
\tl X_i=X_i+\xi_i^\al(0,x) \la_\al; \quad \la \in \HH^*,\
\lug \la ,[\HH,\HH]\rug=0,
$$
are nontrivial. It can also be shown that there are no other nontrivial
continuations.

We now outline the main points of the proof of Theorem 1.

{\bf Proof of Theorem 1.}
Using (2.4), we write Eq. (2.2) in more detail as
$$
\al_i^a(q)\p_{q^a}\al_j^b(q)-\al_j^a(q)\p_{q^a}\al_i^b(q)=C_{ij}^k\al_k^b(q);
\eq (2.6)
$$
$$
\al_i^a(q)\p_{q^a}\chi_j(q;\la)-\al_j^a(q)\p_{q^a}\chi_i(q;\la)=C_{ij}^k
\chi_k(q;\la). \eq (2.7)
$$
Equation (2.6) is equivalent to $[a_i,a_j]=C_{ij}^ka_k$, where
$a_i=\al_i^a(q)\p_{q^a}$,
and the operators $a_i$, are therefore generators of the transformation
group acting in the domain $Q$ ($Q = G/H$ for a real polarization and
$Q = G^c/H$ for a complex polarization, where $H$ is the Lie group
corresponding to $\HH$). Because $\rank \al^a_i(q)=\dim \Ol/2=\dim Q$,
we conclude that whenever solutions of
system (2.6) exist, there exists an isotropy algebra $\HH$ of the point
$q = 0$ of the dimension $n - \dim\Ol/2$. It is obvious that the converse
is also true: the existence of an $(n-\dim Q)$-dimensional subalgebra
$\HH$ is sufficient for the existence of solutions to system (2.6).

As follows from Proposition 2, the existence of solutions to system (2.7)
with initial conditions (2.1)
(or the existence of nontrivial continuations of the operators $a_i$) is
equivalent to the condition that the algebra $\HH$ be adapted to the linear
functional $\la$. Therefore, the existence of the polarization $\HH$
for a covector $\la$ is necessary and sufficient for relations (2.1)
and (2.2) to be satisfied.

We now show that the normality of the polarization, i.e., that
the polarization satisfies the {\it Pukanski conditions}
$\la + \HH^\bot\subset \Ol$, is
necessary and sufficient for relations (2.3) to be satisfied. We
let $\{e_A\}$ denote a basis of the isotropy algebra $\HH$ and
$\{e_a\}$ denote the complementary
basis vectors in $\G^c$: $e_i=\{e_A,e_a\}$. By definition of the
isotropy algebra, $\al_A^a(0)=0$, whence $\det \al_a^b(0)\neq 0$.
Making a linear change of coordinates $q$
and of the basis in $\G^c$, we can assume without loss of generality that
$\al_a^b(0)=\delta_a^b$. In our notation, $\HH^\bot=\{(0,p_a)\}$, where
$p_a$ are arbitrary numbers,
and the Pukanski condition becomes $(\la_A,p_a+\la_a)\in \Ol$.

Let relations (2.3) be satisfied. Setting $q = 0$ in (2.3), we then have
$$
\Phi(\la_A,\la_a)=\Phi(f_A(q,p;\la),f_a(q,p;\la))\bigr|_{q=0}=
\Phi(\la_A,p_a+\la_a).
$$
(We recall the notation $\Phi=(F^\s,K^\s)$).
This implies that for any value of $p_a$, the point
$(\la_A,p_a+\la_a)$ belongs to
the same class of orbits as the point $(\la_A,\la_a)$; because the class
of orbits consists of adenumerable number of K-orbits, we then conclude
that these two points belong to the same orbit.

Conversely, let the Pukanski condition
be satisfied, which means that
$f(0,p;\la)=(\la_A,p_a+\la_a)\in \Ol$ and
$\Phi(\la_A,p_a+\la_a)=\Phi(\la_A,\la_a)$.
We then show that Eq. (2.3) is satisfied. Because the
functions $\Phi(f)$ satisfy Eqs. (1.1) and (1.3), we have
$$
\left(\al_b^a(q)
\frac{\p \Phi(f(q,p;\la))}{\p q^a}-\frac{\p f_b(q,p;\la)}{\p q^a}
\frac{\p \Phi(f(q,p;\la))}{\p p_a}\right)\biggr|_{q=0}=
\frac{\p \Phi(f(q,p;\la))}{\p q^b}\biggr|_{q=0}=0.
$$
Therefore, $\Phi(f(q,p;\la))=\Phi(f(0,p;\la))=\Phi(\la)$, and Theorem 1
is proved.

It is known that a solvable polarization exists for an arbitrary Lie
algebra and any nondegenerate covector. On the other hand, if the algebra
$\G$ is solvable, every functional has a polarization $\HH \subset \G^c$.
In the classical method of orbits, the polarization appears as an
$n-\dim \Ol/2$-dimensional subalgebra $\HH \subset \G^c$, with its
one-dimensional representation determined by the functional $\la$.
In our case, a normal polarization determines linear transition (2.4)
to the canonical coordinates.

It can be easily seen that replacing the functional $\la$ with another
covector belonging
to the same orbit leads to replacing the polarization $\HH$
with the conjugate one $\tl \HH$, with the Darboux coordinates
corresponding to these two polarizations related by a point transformation,
$\tl q^a=\tl q^a(q);\ \tl p_a=\frac{\p q^b}{\p \tl q^a}p_b$.
Therefore, the choice of a specific representative of the
orbit is not essential. On the other hand, if the polarizations are not
conjugate, the corresponding Darboux coordinates are related by a more
general canonical transformation. With the "quantum" canonical
transformation determined (with $q$ and $p$ being operators, see below),
we can thus construct the intertwining operator between the two
representations obtained via the method of orbits involving two
polarizations.

In the case where no polarization exists for a given
functional, the transition to Darboux coordinates (which is nonlinear in
the $p$ variables) can still be constructed, and the
$\la$-representation of $\G$
can still be defined (see below); this representation is the basis for the
harmonic analysis on Lie groups and homogeneous spaces (applications of the
method of orbits to harmonic analysis go beyond the scope of this paper and
are not considered here). In other words, the existence of a polarization
is a useful property but is not necessary for the applicability of the
method of orbits.

As already mentioned, the functional $\la$ can have several
different polarizations; however, it is easy to verify the following
proposition.

{\bf Proposition 3.} {\it If the normal polarization $\HH$ exists
for a given $\la\in \G^*$, then} $\G^\la \subset \HH$.

{\bf Example 4} (The group $SO(2,1)$, continuation of Example 3).
The Kirillov form on nondegenerate orbits is given by
$\omega_\la=df_2\wedge d f_3/2f_1$.
For different spectral types, we obtain
\be
{\rm type}&A:& f_1=p,\quad f_2=e^q(-p+j),\quad f_3=e^{-q}(p+j);\\
&& \la=(0,j,j);\quad \HH=\{e_2+e_3,e_1+e_2\};\quad
(p,q)\in R^2;\\
{\rm type}&B:& f_1=p,\quad f_2=e^q(-ip+j),\quad f_3=-e^{-q}(ip+j);\\
&& \la=(0,j,-j);\quad \HH=\{e_2-e_3,e_1-ie_2\}.
\ee
To find the domain of definition
of the variables $(p, q)$ for type-B orbits, we decompose the complex
variable $q$ into its real and imaginary components, $q=\al +i\beta$.
Because the variables $f_i$ are real, we obtain
$$
p=j\tan \beta;\quad f_1=j\tan \beta,\quad f_2=\frac{j e^\al}{\cos\beta},
\quad f_3=-\frac{e^{-\al}}{\cos\beta}.
$$
which implies that $p$ takes any real value
and $q$ is a complex variable with its real part in $R^1$ and the
imaginary part in $S^1$, i.e., $q\approx q + 2\pi i$. Therefore,
functions on a type-B K-orbit are
analytic functions of a real variable $p$ and a complex variable $q$
that are $2\pi i$-periodic in $q$.

We define the notion of the quantization of K-orbits.
This quantization is to be done separately for each spectral type of
orbits; it consists in assigning the spectral type of the orbit a special
representation of the Lie algebra (the $\la$-representation),
with the orbits
subject to the {\it integrai-veduedaess} condition considered in the next
section.

We now view the transition functions $f_i(q,p;\la(j))$ to local
canonical coordinates as symbols of operators that are defined as follows:
the variables $p_a$ are replaced with derivatives,
$p_a\to \hat p_a\equiv -i\hbar \frac{\p}{\p q^a}$, and
the coordinates of a covector $f_i$ become the linear operators
$$
f_i(q,p;\la(j))\to \hat f_i\left(q,-i\hbar
\frac{\p}{\p q};\la(\tilde j)\right)
$$
(with $\h$ being a positive real parameter). This quantization
procedure is obviously ambiguous. This ambiguity is eliminated if we
require that the operators $\hat f_i$ satisfy the commutation relations
$$
\frac{i}{\hbar}[\hat f_i,\hat f_j]=C_{ij}^k\hat f_k. \eq (2.8)
$$
If the transition to the canonical coordinates is linear, i.e., a
normal polarization exists for orbits of a given type, it is obvious that
$$
\hat f_i=-i\hbar \al_i^a(q)\frac{\p}{\p q^a}+\chi_i(q,\la(\tilde j)),
$$
and Eq. (2.8) is equivalent to conditions
(2.6) and (2.7).

Under quantization, an arbitrary analytic function $\varphi(f)$
on the coalgebra is mapped into a symmetrized operator function
$\varphi(\hat f)$ of
the operators $\hat f_i$. The parameters $j$ are related to the
parameters $\tilde j$ of the
orbit as $\tilde j=j+i\hbar \beta$, where $\beta$ is an $\rs$-dimensional
real vector that is to
be found from the condition that the functions
$$
\kappa^\s_\mu(j)=\Ksm(\hat f). \eq (2.9)
$$
are real. We note that in the "classical" limit as $\h \to 0$,
we have $\kappa^\s_\mu(j) \to \os_\mu(j)$, and
the commutator of linear operators goes into the Poisson bracket on the
coalgebra,
$$
\frac{i}{\hbar}[\cdot,\cdot]\to \{\cdot,\cdot\}.
$$

Because the functions $\kappa^\s_\mu(j)$ are generally different
from the functions $\os_\mu(j)$ for $\h\neq 0$, we must redefine the domain
of definition $J^\s$ of the $j$ parameters such that each point $j\in J^\s$
 is in a one-to-one correspondence with the values of $\kappa^\s_\mu(j)$,
i.e., the functions $(\kappa^\s)^{-1}$ are
unambiguous upon restrictions to $J^\s$. The condition
$$
\kappa^\s(J^\s)=\Omega^\s. \eq (2.10)
$$
must also be satisfied.

We introduce the operators
$$
l_k(q,\p_q,j)\equiv \frac{i}{\hbar}\hat f_k(q,\hat p;\la(\tilde j)).
$$
It is obvious that
$$
[l_i,l_j]=C_{ij}^kl_k;\quad \Fsa(-i\hbar l(q,\p_q,j))\equiv 0,\
\Ksm(-i\hbar l(q,\p_q,j))\equiv \kappa^\s_\mu(j);
$$
$$
\kappa^\s_\mu(j)=\overline{\kappa^\s_\mu(j)};\quad
\det \frac{\p \kappa^\s_\mu(j)}{\p j_\nu}\neq 0;\quad j\in J,\ q\in Q.
\eq (2.11)
$$

{\bf Definition 1.} Let $f_i=f_i(q,p;\la(j))$ be a transition to canonical
coordinates on the orbit ${\cal O}_{\la(j)}$ and $\la(j)$ be a parameterized
covector. The realization of the Lie algebra $\G$ by the respective
operators $l_i(q,\p_q,j)$ is called the $\la$-{\it representation}.

In what follows, we show that the quantities $\kappa^0_\mu(j)$ constitute
the spectrum of Casimir operators on the Lie
group $\hat K_\mu\equiv K_\mu(i\hbar \xi(g))$, where $\xi_i(g)$
are left-invariant vector fields on $G$.
Accordingly, $\kappa^\s_\mu(j)$ are the eigenvalues of the Casimir operators
$\hat \Ksm\equiv \Ksm (i\hbar X)$, where $X_i$ are the transformation group
generators, on the homogeneous $(s)$-type space [5].

{\bf Example 5} (The group $SO (2, 1)$, continuation of Example 3).
\footnote{Here and in what follows, the symbol $"\circ"$ denotes the
symmetrized product of operators, $A\circ B\equiv \frac12 (AB+BA)$.}
$$
{\rm Type}\ A:\ \hat f_1=-i\h\frac{\p}{\p q},\ \hat f_2=e^q(i\h\frac{\p}
{\p q}+j+i\h\beta),\ \hat f_3=e^{-q}(-i\h\frac{\p}{\p q}+j+i\h\beta);
$$
$$
\kappa(j)=K(\hat f)=\hat f_1^2+\hat f_2\circ\hat f_3=
j^2+\h^2\beta(1-\beta)+i\h j(2\beta-1).
$$
Because $\kappa(j)$ is real, we
obtain $\beta=\frac12$. Then the $\la$-representation for spectral type A
is given by
$$
l_1=\frac{\p}{\p q},\ l_2=e^q(-\frac{\p}{\p q}+\frac{i}{\h}j-\frac12),\
l_3=e^{-q}(\frac{\p}{\p q}+\frac{i}{\h}j-\frac12);\quad
\kappa(j)=j^2+\frac{\h^2}{4},
$$
where $j\in J^A=[0,\infty)$ and
$\Omega^A\equiv \kappa(J^A)=[\frac{\h^2}{4},\infty)$
$$
{\rm Type}\ B:\ \hat f_1=-i\h\frac{\p}{\p q},\
\hat f_2=e^q(-\h\frac{\p}{\p q}+j+i\h\beta),\
\hat f_3=e^{-q}(-\h\frac{\p}{\p q}-j-i\h\beta);
$$
$$
\kappa(j)=K(\hat f)=\hat f_1^2+\hat f_2\circ\hat f_3=
-j^2+\h^2\beta^2-\h j-i\h \beta(2j+\h).
$$
Because $\kappa(j)$ is real, we obtain $\beta=0$; the $\la$-representation
of spectral type B is given by
$$
l_1=\frac{\p}{\p q},\ l_2=e^q(-i\frac{\p}{\p q}+\frac{i}{\h}j),\
l_3=e^{-q}(-i\frac{\p}{\p q}-\frac{i}{\h}j);\quad \kappa(j)=-j(j+\h),
$$
where $j\in J^B=[-\frac{\h}{2},\infty)$ and
$\Omega^B\equiv \kappa(J^B)=(-\infty,\frac{\h^2}{4})$.
Condition (2.10) is satisfied, $\Omega^A\cup \Omega^B=\Omega=R^1$.

{\bf Example 6} (The group $St(1,R)$). Kirillov [3] gives the group
$St(1,R)$ as an example
illustrating the absence of the polarization for a general covector. We
describe the structure of K-orbits of this group. The corresponding Lie
algebra can be realized by the matrices
$$
X(a,\xi,c)=
\pmatrix{
0&\xi_1&\xi_2&c\cr
0&a_1&a_2&\xi_2\cr
0&a_3&-a_1&-\xi_1\cr
0&0&0&0\cr
} \equiv a_1e_1+a_2e_2+a_3e_3+\xi_1e_4+\xi_2e_5+ce_6.
$$
The basis elements $e_i$,
have the nonvanishing commutation relations
\be
&&[e_1,e_2]=2e_2,\quad [e_1,e_3]=-2e_3,\quad [e_1,e_4]=-e_4,\quad
[e_1,e_5]=e_5,\\
&&[e_2,e_3]=e_1,\quad [e_2,e_4]=-e_5,\quad [e_3,e_5]=-e_4,\quad
[e_4,e_5]=2e_6.
\ee
We now describe orbit (1.4) for this group as
\be
O^0_\omega&=&\{K_1(f)=\omega^0_1,\ K_2(f)\equiv f_6=\omega^0_2; \
\neg(F^1(f)=0)\};\\
O^{1a}_\omega&=&\{ F^{1a}_1(f)=F^{1a}_2(f)=F^{1a}_3(f)=0,\quad
f_6=\omega^{1a}\neq 0\};\\
O^{1b}_\omega&=&\{ f\neq 0; \quad f_4=f_5=f_6=0,\quad
f_2f_3+f_1^2/4=\omega^{1b}\};\\
O^2&=&\{f=0\};\
K_1(f)\equiv f_1^2f_6-f_1f_4f_5+f_4^2f_2-f_5^2f_3+4f_2f_3f_6;\\
F^{1a}_1(f)&\equiv &2f_1f_6-f_4f_5;\  F^{1a}_2(f)\equiv 4f_2f_6-f_5^2;\
F^{1a}_3(f)\equiv 4f_3f_6+f_4^2.
\ee
It can be easily seen that the polarization does not exist for the
degenerate orbits $O^{1a}_\omega$. The Kirillov form on orbits of this
type is given by $\omega_\la=df_4\wedge d f_5/2f_6$.
The space $M_{1a}$ is of the first spectral type, and we can
introduce global coordinates on $B_{1a}$, i.e., a parameterization
$\la(j)=(0,0,0,0,0,j)$. For this functional, we have
$\G^\la=\{e_1,e_2,e_3,e_6\},\ {\cal N}_\la=\{e_1,e_2,e_3\}$, and
${\cal K}_\la=\{e_6\}$. For the orbits under consideration, linear
transition (2.4) to canonical coordinates is replaced by
$$
f_1=qp,\ f_2=q^2j,\ f_3=-p^2/4j,\ f_4=p,\ f_5=2qj,
\ f_6=j;\quad (p,q)\in R^2,
$$
which is quadratic in the $p$ variables. We now construct the corresponding
$\la$-representation. We do not encounter problems with ordering the
operators $\hat p=-i\h\p_q$ and $\hat q=q$ involved in the operators
$\hat f_2=q^2j,\; \hat f_3=-{\hat p}^2/4j,\; \hat f_4=\hat p,\;
\hat f_5=2qj,\; \hat f_6=j$
The operator $\hat f_1$ is unambiguously found from the
commutation relations as $\hat f_1=q\hat p-i\h/2$. For orbits of this
type, the $\la$-representation is therefore given by
$$
l_1=q\p_q+\frac12,\ l_2=\frac{i}{\h}q^2j,\ l_3=\frac{i\h}{4j}\p_q^2,\
l_4=\p_q,\ l_5=\frac{i}{\h}2qj,\ l_6=\frac{i}{\h}j;\
\kappa^{1a}(j)=j\in R^1.
$$

\section{The integral-valuedness condition
for the orbits and the spectra of Casimir operators}

In this section, we
describe the first stage in the explicit construction of harmonic analysis
on homogeneous spaces. This is an involved subject, however, worthy of a
separate investigation, and we practically do not consider it here. In this
section, we show that the functions $\kappa(j)$ introduced above are the
eigenvalues of the Casimir operators, with the parameters $j$ satisfying the
integral-valuedness condition.

On a connected and simply-connected real Lie
group $G$, we introduce the quasi-invariant measure
$dg=\sqrt{d_lgd_rg}$, where $d_lg$
and $d_rg$ are the left and right Haar measures. In the space
$L_2(G,dg)$, we define a unitary representation of $G\times G$ via
$$
T_{(g_1,g_2)}u(g)=\sqrt{\frac{d(g_1^{-1}gg_2)}{dg}}u(g_1^{-1}gg_2),\quad
u(g)\in L_2(G,dg). \eq (3.1)
$$
The infinitesimal generators of representation (3.1) are the left-and
right-invariant operators $\xi_i$ and $\eta_i$ given by
$$
\xi_i(g)=\xi_i^j(g)\p_{g^j}+C_i,\quad
\eta_i(g)=\eta_i^j(g)\p_{g^j}+C_i;\quad C_i\equiv-\frac14 C_{ij}^j.
$$
The generators $\xi_i$ and $\eta_i$, differ by additive constants from
(for a unimodular group, coincide
with) the corresponding left- and right-invariant vector fields on the
group $G$, and we also refer to them as vector fields.

Because the
eigenfunctions of unbounded operators with continuous spectra do not belong
to $L_2(G,dg)$ and are instead linear functionals on a dense set (a nuclear
space) $\Phi\subset L_2(G,dg)$, we must consider the Gelfand triplet
$\Phi\subset L_2(G,dg)\subset \Phi'$,
where $\Phi'$ is the space dual to $\Phi$.

Similarly to (1.4), we decompose the space $L_2(G,dg)$ into subspaces
that are invariant with respect to
representation (3.1) as $L_2(G,dg)=\cup_\s\LL_\s$, where
$$
\LL_\s=\{\vp(g)\in L_2(G,dg) \mid \Fsa(\xi)\vp(g)=0,\
\neg(F^{s+1}(\xi)\vp(g)=0)\}. \eq (3.2)
$$
We note that replacing the left-invariant
fields $\xi_i$ in (3.2) with the right-invariant $\eta_i$ does not
change the spaces $\LL_\s$. (This can be seen, for example, using
the mapping $g \to g^{-1}$, under
which the left-invariant fields pass into right-invariant ones and vice
versa, and the measure $dg$ remains invariant.) For each $\LLs$, we also
introduce the Gelfand triplet $\Phi_\s\subset \LLs\subset \Phi'_\s$.

On each space $\LL_\s$, there
exist bi-invariant Casimir operators $\Ksm(i\h \xi)(=\Ksm(-i\h\eta))$,
and the space $\LL_\s$ can be decomposed into a direct sum
(a direct integral) of
eigensubspaces of the Casimir operators, i.e., we observe complete
similarity with decomposition (1.4).

{\bf Theorem 2.} {\it Let $\la(j)$ be a parameterized
$(s)$-covector. The quantities $\kappa^\s_\mu(j)$, Eq. (2.9), are
eigenvalues of the Casimir operators $\Ksm(i\h\xi)$ on $\LL_\s$,
where the parameters $j$ satisfy the condition
$$
\lug \la(j),e_\mu\rug =2\pi\h n_\mu/T_\mu; \quad n_\mu \in Z, \eq (3.3)
$$
where $e_\mu$ is the basis vector of the
one-dimensional Lie algebra of the one-parameter compact subgroup of the
commutative quotient group $K_\la$ and $T_\mu$ the period of the
one-dimensional compact subgroup} ($\exp(T_\mu e_\mu)=1$).

To find the spectra of the Casimir
operators, it is therefore sufficient to find the functions
$\kappa^\s_\mu(j)$ (these
functions are actually given by the structure constants and determine the
spectra of the Casimir operators on the universal covering group $\tl G$)
and impose quantization condition (3.3) on the parameters $j$. We shortly
demonstrate that condition (3.3) is equivalent to the Kirillov
{\it integral-valuedness} condition for K-orbits.

{\bf Proof of Theorem 2.} We discuss
the main points of the proof. Let $\la(j)$ be a parameterized
$(s)$-covector and $l_i(q,\p_q,j)$
be the corresponding $\la$-representation. We define the
distributions $D_q^j(g)$ on $G$ by the equations
$$
[\xi_i(g)+l_i(q,\p_q,j)] D_q^j(g)=0, \quad i=1,\dots,n. \eq (3.4)
$$
From the definition of $\la$-representation (2.11), it is easy to obtain
that
$$
\Fsa(\xi)D_q^j(g)=0,\quad {\rm i.e.,}\quad D_q^j(g)\in \Phi'_\s;\quad
\Ksm(i\h\xi)D_q^j(g)=\kappa^\s_\mu(j)D_q^j(g).
$$
Although system (3.4) is compatible, the functions
$D_q^j(g)$ do not exist globally on
the entire group for all values of the $j$ parameters. We restrict system
(3.4) to the subgroup $G^\la$ and set $q = 0$. Recalling Proposition 1,
we then obtain
$$
[\xi_A(g_\la)+\frac{i}{\h}\la_A(\tl j)]D^j_0(g_\la)=0,\quad
g_\la\in G^\la, \eq (3.5)
$$
where the
subscript $A$ labels the basis vectors of $\G^\la$. It follows from
(3.5) that the functions $D_q^j(g)$ are globally defined only if the
condition
$$
\frac1{2\pi \h}\oint\limits_{\gamma\in H_1(G_\la)}\omega^j_{G^\la}=n_\gamma \in Z.
\eq (3.6)
$$
is satisfied, where $\omega^j_{G^\la}=\omega^A \la_A(j)$
is a closed left-invariant 1-form on the group $G^\la$. (We have thus
rediscovered the Kostant criterion [8].) Condition (3.6) is equivalent to
the Kirillov {\it itegral-valuedness} condition for orbits [3].
Using the
results in the previous sections, we can represent integral-valuedness
condition (3.6) in a more detailed form. For a chosen parameterized
covector $\la(j)$, the basis of left-invariant vector fields of $\G^\la$
is spanned by the vectors ($\tl \la\equiv \la(\tl j)$)
\be
\nabla \Fsa(g)&\equiv&(L_g)_*\nabla \Fsa(\tilde \la)=
F^{\s i}_\al(\tilde\la)\xi_i(g);\\
\nabla \Ksm(g)&\equiv&(L_g)_*\nabla \Ksm(\tilde \la)=
K^{\s i}_\mu(\tilde\la)\xi_i(g),
\ee
where
$$
F^{\s i}_\al(\tilde\la)=\frac{\p \Fsa(f)}{\p f_i}
\biggr|_{f=\tilde \la},\quad
K^{\s i}_\mu(\tilde\la)=\frac{\p \Ksm(f)}{\p f_i}\biggr|_{f=\tilde \la},
$$
and $L_g$ is the left-invariant
representation of $G \times G$ on $L_2(G,dg)$. We rewrite system (3.5) as
$$
\nabla \Fsa(g_\la)D^j_0(g_\la)=0; \quad g_\la \in G_\la; \eq (3.7)
$$
$$
[\nabla \Ksm(g_\la) + \frac{i}{\h}m^\s_\mu\omega^\s_\mu(\tilde j)]
D^j_0(g_\la)=0; \quad g_\la \in G^\la. \eq (3.8)
$$
The operators $\nabla \Fsa$ entering Eq. (3.7)
are left-invariant vector fields on the normal subgroup $N_\la$, and the
function $D^j_0(g_\la)$ is therefore independent of the coordinates of
the element $n$
entering the decomposition $g_\la=nk;\ n\in N_\la,\ k\in K_\la$. Similarly,
the operators $\nabla \Ksm$ are left-invariant vector fields on the
commutative group $K_\la$, and
system (3.8) is therefore an eigenvalue problem for $\rs$ commuting first-
order operators. This problem is easily solved. It is obvious that if the
one-parameter subgroup with the generator $\nabla \Ksm$ is noncompact, the
corresponding eigenvalue $m^\s_\mu\omega^\s_\mu(\tilde j)$ can take any
value, i.e., the parameters
$j$ are not quantized in that case. With a basis chosen to be independent of
the parameters $j$ (such a basis always exists because the Casimir functions
are homogeneous), system (3.8) for compact subgroups of $K_\la$ becomes
$$
[\frac{\p}{\p k^\mu} + \frac{i}{\h}\la_\mu(j)]D^j_0(k)=0.
$$
Therefore, the integral-valuedness condition is given by Eq. (3.3).

We have thus shown that the functions $\kappa^\s_\mu(j)$ of the integer
parameters $j$ belong to the spectrum of Casimir operators on $\LL_\s$.
To complete the proof, it is necessary to show that the family of functions
$D^j_q(g)$ is dense in $\Phi'_\s$, which in turn implies the absence of
other elements of the
spectrum. There is convincing evidence that this is the case [5], but the
discussion of this point takes us far beyond the subject of the present
paper.

{\bf Example 7} (Group SO(2, 1), continuation of Example 3). For spectral
type $A$, we have $\la(j)=(0,j,j)$ and $\G^\la=\{ e_2+e_3\}$;
for spectral type $B$, $\la(j)=(0,j,-j)$ and $\G^\la=\{ e_2-e_3\}$.
Therefore, to single out integral
orbits, we must know whether the corresponding one-parameter groups are
compact. It is obvious that for spectral type $A$, the group
$K_\la=\exp [(1/2) t(e_2+e_3)]$
is noncompact, and all orbits of this type are
therefore integral, i.e., the parameter $j$ is not quantized;
$\kappa(j)=j^2+\h^2/4,\ j\geq 0$.
For spectral type $B$, the group
$K_\la=\exp [(1/2) t(e_2-e_3)]$ is compact
and has the period $T=2\pi$. From Eq. (3.3), we see that
$\lug \la(j),(1/2) (e_2-e_3)\rug=j=n\h$,
whence $j=n\h,\ n\in Z$ end $\kappa(j)=-\h^2 n(n+1)$ for $n=0,1,\dots.$

\section*{REFERENCES}
\begin{enumerate}
\item A. A. Kirillov, {\sl Rus. Math. Surv.,} {\bf 17,} 53--104 (1962).
\item A. A. Kirillov, {\sl Funct. Anal. Appl}, {\bf 2,} 133--146 (1968);
{\bf 3,} 29--38 (1969).
\item A. A. Kirillov, {\sl Elements of the Theory of Representations}
[in Russian], Nauka, Moscow (1978); English transl. prev. ed.,
Springer, Berlin (1975).
\item A. A. Kirillov, "Introduction to representation theory and
noncommutative harmonic analysis [in Russian]," in:
{\sl Representation Theory and Noncommutative Harmonic Analysis 1}
(Results of Science and Technology: Problems in Mathematics:
Fundamental Directions, Vol. 22) (R. V. Garnkrelidze, ed.), VINITI,
Moscow (1988), pp. 5--162.
\item I. V. Shirokov, "K-orbits, harmonic analysis on homogeneous spaces,
and integrating differential equations [in Russian]," Preprint,
Omsk State Univ,.Omsk (1998).
\item A. V. Shapovalov and I. V. Shirokov, {\sl Theor. Math. Phys.},
{\bf 104,} 921--934 (1995).
\item A. S. Mishchenko and A. T. Fomenko, {\sl Selecta Math. Soviet},
{\bf 2,} 207--291 (1982).
\item B. Kostant, "Quantization and unitary
representations: I. Prequantization," in: {\sl Lectures in Modern
Analysis and Applications III} (Lect. Notes Math., Vol. 170), Springer,
Berlin (1970), pp. 87-208.
\end{enumerate}
\end{document}